\documentclass[aps,twocolumn,nofootinbib,preprintnumbers,citeautoscript]{revtex4}
\usepackage{amsmath,graphicx}

\def\be{\begin{equation}}
\def\ee{\end{equation}}

\def\n{\nu}
\def\a{\alpha}
\def\e{\epsilon}
\def\k{\kappa}
\def\b{\beta}
\def\d{\delta}

\def\D{\Delta}
\def\G{\Gamma}
\def\l{\lambda}

\def\G{\Gamma}

\begin{document}
\title{Conformal Bootstrap in Mellin Space}
\smallskip\smallskip %%\smallskip
\author{Rajesh Gopakumar$^{1}$\thanks{email: rajesh.gopakumar@icts.res.in},
Apratim Kaviraj$^{2}$\thanks{email: apratim@cts.iisc.ernet.in},  Kallol Sen$^{2,3}$\thanks{email: kallolmax@gmail.com} and
{Aninda Sinha}$^{2}$\thanks{email: asinha@cts.iisc.ernet.in}
\\[2mm]
\it $^1$International Centre for Theoretical Sciences (ICTS-TIFR), 
%\\\it Survey No. 151, 
Shivakote, Hesaraghatta Hobli,
Bangalore 560089, India  \\[2mm]
\it $^2$Centre for High Energy Physics, Indian Institute of
Science, %\\\it 
C. V. Raman Avenue, Bangalore 560012, India\\[2mm]
\it $^3$Kavli Institute for the Physics and Mathematics of the Universe (WPI),The University of Tokyo Institutes for Advanced Study, The University of Tokyo, Kashiwa, Chiba 277-8583, Japan}

%%\vskip2cm \abstract{ 
\begin{abstract}
{We propose a new approach towards analytically solving for the dynamical content of Conformal Field Theories (CFTs) using the bootstrap philosophy. This combines the original bootstrap idea of Polyakov with the modern technology of the Mellin representation of CFT amplitudes. We employ exchange Witten diagrams with built in crossing symmetry as our basic building blocks rather than the conventional conformal blocks in a particular channel. Demanding consistency with the operator product expansion (OPE) implies an infinite set of constraints on operator dimensions and OPE coefficients. We illustrate the power of this method in the epsilon expansion of the Wilson-Fisher fixed point by reproducing anomalous dimensions and, strikingly,  obtaining OPE coefficients to higher orders in epsilon than currently available using other analytic techniques (including Feynman diagram calculations).   
Our results enable us to get a somewhat better agreement of certain observables in the 3d Ising model, with the precise numerical values that have been recently obtained.  }
\end{abstract}
  
\maketitle

\newpage
\section{Introduction}
\label{sec:introduction}

The Wilsonian paradigm \cite{wilsonfisher, wilson, wilsonkogut} for quantum field theories centre-stages the scale invariant fixed points of the renormalisation group. In the context of relativistic QFTs these critical points are believed to be conformally invariant \cite{Polyakov:1970xd}. The study of such conformal field theories (CFTs) is thus central to many areas of physics. Unfortunately, we currently have very few tools to access the dynamics of such CFTs, apart from cases where they are free or close to free.  The dynamical data of a CFT is entirely in its spectrum of dimensions of primary operators as well as their three point functions (or OPE coefficients). In principle, conformal invariance and associativity of the Operator Product Expansion (OPE) in the four point function give powerful constraints on this data \cite{fgg,Polyakov}. In practice, apart from two dimensions \cite{bpz}, this constraint has been difficult to effectively implement. 

Recently, there has been a successful revival \cite{rrtv, review} of this bootstrap program, in which associativity and positivity constraints have been translated into inequalities which can be efficiently implemented numerically using linear programming \cite{rrtv}, semi-definite programming \cite{mostprecise} and judicious truncation \cite{gliozzi}. This has led to rather amazing constraints on the low lying  spectrum (as well as OPE coefficients) of various nontrivial CFTs -- see \cite{review} for references. These numerical techniques now give the best data on the low lying operators of the 3d Ising model \cite{3dising,mostprecise,susyboot} and hint at there being special points in the domains allowed by the inequalities.
 
Here, we will outline a new approach to the conformal bootstrap for $CFT_d$ which is calculationally effective as well as being conceptually suggestive. This involves two ingredients which turn out to blend very naturally. The first involves revisiting an approach of Polyakov which has crossing symmetry from the outset but is not obviously 
compatible with the operator expansion. We will implement this approach in terms of conformally invariant building blocks which are exchange Witten diagrams in $AdS_{d+1}$  rather than the conventional conformal blocks.  In other words, for a four point function of identical external scalars, we expand the amplitude as a function of cross ratios $(u,v)$ in terms of functions $W_{\Delta,\ell}^{(s)}(u,v)$ which can be written in terms of an integral of $AdS_{d+1}$ bulk to bulk propagator (corresponding to a $CFT_d$ operator of dimension $\D$, spin $\ell$) together with bulk to boundary propagators for the four external scalars of dimension $\Delta_\phi$.
\begin{eqnarray}\label{polyapp}
&&\!\!\!\!\!\!\!{\cal A}(u,v)= \langle {\cal O}(1){\cal O}(2){\cal O}(3){\cal O}(4) \rangle \nonumber \\
&=& \!\!\!\!\!\sum_{\Delta, \ell} \!\!c_{\Delta, \ell}\bigg( W_{\Delta,\ell}^{(s)}(u,v)+W_{\Delta,\ell}^{(t)}(u,v)+W_{\Delta,\ell}^{(u)}(u,v) \bigg).
\end{eqnarray}
The sum here is over the entire physical spectrum of primary operators generically characterised by the dimensions $(\Delta)$ and spin ($\ell$). The coefficients $c_{\Delta,\ell}$ are proportional to (square of) the OPE coefficients. The central observation of Polyakov \cite{Polyakov} was that there are spurious powers, in this case, $u^{\Delta_\phi}$ (and $u^{\Delta_\phi}\ln{u}$) in such an expansion. Demanding cancellations of these terms (as a function of $v$) gives an infinite number of constraints on $\Delta$ as well as the coefficients $c_{\Delta, \ell}$.   

The second ingredient exploits the Mellin representation \cite{mack,pene,mig,aldaybissi,rastelli} of CFT amplitudes which is a close counterpart of momentum space in usual QFTs. For the above amplitude, this representation is essentially a Mellin transform  w.r.t. the cross ratios.  
\be\label{idmelldef}
{\cal A}(u,v)= \int_{-i\infty}^{i\infty}\frac{ds}{2\pi i} \frac{dt}{2\pi i} u^{s}v^t \rho_{\Delta_{\phi}}(s,t)
{\cal M}(s,t) \ . 
\ee
Here $\rho_{\Delta_{\phi}}(s,t)= \Gamma^2(-t)\Gamma^2(s+t)\Gamma^2(\Delta_{\phi}-s)$ is a convenient kinematic factor while ${\cal M}(s,t)$ contains the dynamics. The integral is evaluated by closing the contour appropriately picking up the poles of the integrand. 
The expansion in (\ref{polyapp}) can now be translated into Mellin space. Each of the Witten exchange functions $W_{\Delta,\ell}^{(s)}(u,v) \rightarrow M_{\Delta,\ell}^{(s)}(s,t)$, which will be discussed below. We use these functions as our basis for an expansion of the (reduced) Mellin amplitude \cite{commconv}. 
\be\label{Mexp}
{\cal M}(s,t)=    \sum_{\Delta, \ell} c_{\D, \ell} (M_{\Delta,\ell}^{(s)}(s,t)+M_{\Delta,\ell}^{(t)}(s,t)
+M_{\Delta,\ell}^{(u)}(s,t))\ .
\ee 

The spurious powers  $u^{\Delta_\phi}$ (and $u^{\Delta_\phi}\ln{u}$) in ${\cal A}(u,v)$ translate into spurious single and double poles in the full Mellin amplitude
\be\label{polexp}
\Gamma^2(\D_\phi-s) {\cal M}(s,t) = \frac{q^{(2)}_{tot}(t)}{(s-\Delta_{\phi})^2} +  \frac{q^{(1)}_{tot}(t)}{(s-\Delta_{\phi})} +\cdots
\ee 
The $\cdots$ refer to physical contributions as well as spurious descendant poles.
Compatibility with the operator expansion demands that we set both residues 
\be\label{qsum}
q^{(a)}_{tot}(t)=\sum_ {\Delta, \ell} c_{\D, \ell} (q^{(a,s)}_{\D, \ell}(t)+q^{(a,t)}_{\D, \ell}(t)+q^{(a,u)}_{\D, \ell}(t) )=0
\ee
for $(a=1,2)$. The terms on the RHS come from the obvious expansion of individual terms in (\ref{Mexp}) in terms of poles as in (\ref{polexp}). This is our central constraint equation. 

We will see that this scheme is calculationally effective by revisiting the $\epsilon$ expansion in 
$(d=4-\e)$ dimensions for a single real scalar at the Wilson-Fisher fixed point. We will find that we can reproduce the answers \cite{wilson, wilsonkogut} for the dimensions of $\phi$ and $\phi^2$ to $O(\e^3)$ and $O(\e^2)$ respectively. For the higher spin currents $J^{(\ell)}$ of the schematic form $\phi\partial^{\ell}\phi$, we reproduce the known anomalous dimensions to $O(\e^2)$ \cite{wilsonkogut} as well as the $O(\e^3)$ piece \cite{gracey}. More nontrivially, we also determine OPE coefficients which are usually difficult to compute using Feynman diagram techniques. Thus we find, for the first time, the three point function $C_{\ell}$ of two $\phi$'s with $J^{(\ell)}$ to $O(\e^3)$. In particular, this enables one to compute the central charge $c_{T}$ which is related to the stress tensor coefficient $C_{\ell=2}$ to this order. Similarly, we will also indicate how to reproduce and go beyond some of the existing results for large spin operators, which were obtained with the (double) light cone expansion \cite{dlce,others,alday}. 

\section{Implementing Mellin Bootstrap}
\label{sec:implement}

Witten exchange functions $W_{\Delta,\ell}^{(s)}(u,v)$ are computed from a tree level four point function with the exchange of a field in $AdS_{d+1}$ of spin $\ell$ (and corresponding to a conformal dimension $\Delta$ on the boundary) in the $s$-channel \cite{witten}. By construction they preserve all the isometries of $AdS_{d+1}$ and  are conformally covariant. Their expressions are unfortunately quite complicated in position space \cite{poswit}. As has been stressed in the literature, there is dramatic simplification in Mellin space. Thus for a scalar exchange 
$M_{\Delta,\ell =0}^{(s)}(s,t)$ can be written in terms of  a ${}_3F_2$ hypergeometric function (evaluated at unit argument). See, for instance, \cite{mig, pene}. 
%\begin{align}
%\begin{split}
%&M_{\Delta,\ell =0}^{(s)}(s,t) = \frac{1}{2s-\Delta}\frac{\Gamma^2(\Delta_{\phi}+\frac{\Delta-2h}{2})}{\Gamma(1+\Delta-h)} \\
%& {} _3F_2[\{1-\Delta_{\phi}+\frac{\Delta}{2}, 1-\Delta_{\phi}+\frac{\Delta}{2}, \frac{\Delta}{2}-s \}, 
%\{1+\frac{\Delta}{2}-s, 1+\Delta-h\}] \,,
%\end{split}
%\end{align}
%where the ${}_3F_2$ is evaluated at unit argument.
It is a meromorphic function (only of $s$ in this case) which has simple (physical) poles at $2s=\Delta +2m$ where $m=0,1,2\ldots$. 
%Here and below $h=\frac{d}{2}$. 

%{\bf AS: The following paragraph probably can be shortened a bit in the event we need to free up more space.}
It is more generally true that $M_{\Delta,\ell}^{(s)}(s,t)$ is the sum of a meromorphic function with poles at $2s=\Delta -\ell +2m$ plus an additional polynomial in $(s,t)$ of degree at most $\ell-1$. Thus our building blocks are polynomially bounded in Mellin space unlike the conformal blocks which have an exponential behaviour \cite{fk,long}. 
This is what makes them a better choice for a basis to expand in terms of. Moreover, they exhibit the right factorisation on the physical poles in having the same residues as the conformal blocks. The way the Witten exchange functions differ from the conformal blocks is that unlike the latter, they additionally contain the contribution of so-called ``double trace" operators. These are operators of dimension for e.g. $(\Delta_1+\Delta_2)$  i.e. $2\D_{\phi}$ in our case of identical scalars. In a large 
$N$ CFT there are indeed physical operators with this dimension (``two particle states") with $\frac{1}{N}$ corrections. But in a generic CFT this is not the case and ``double trace" operators is a misnomer for these contributions \cite{compl}. They are really spurious contributions which need to cancel out in the full amplitude -- both in position space as well as Mellin space, as discussed above.

Many of these properties of Witten diagrams are transparent in a spectral (or ``split") representation of these diagrams \cite{pene,mig,spinning}. In position space, this can be used to write the Witten functions as 
\be\label{hf}
W_{\Delta,\ell}^{(s)}(u,v) = \int_{-i\infty}^{i\infty} d\nu\, \mu_{\Delta, \ell}(\nu) F^{(s)}_{\nu,\ell}(u,v)\,,
\ee
 where the conformal partial waves $F^{(s)}_{\nu,\ell}(u,v)$ are purely kinematic in nature - their exact form can be found in e.g. \cite{dolanosborn} and will not be important in the following. The spectral function for identical external scalars given by $2\pi i \mu_{\Delta,\ell}(\nu)=\xi_{\Delta,\ell}(\nu)\xi_{\Delta,\ell}(-\nu)$ with 
\be\label{specunitry}
\xi_{\Delta,\ell}(\nu)=\frac{\G^2(\frac{2\D_{\phi}-h+\ell+\nu}{2})}{((\D-h)+\nu)\G(\nu)(h+\nu-1)_\ell}\,,
\ee 
% \be\label{specunitry}
%\!\!\mu_{\Delta,\ell}(\nu)=\frac{\G^2(\frac{2\D_{\phi}-h+\ell+\nu}{2})\G^2(\frac{2\D_{\phi}-h+\ell-\nu}{2})}{2\pi i((\D-h)^2-\nu^2)\G(\nu)\G(-\nu)(h+\nu-1)_\ell(h-\nu-1)_\ell}\,,
%\ee
contains the information about the exchanged operators. The poles (in $\nu$) are at the physical value (together with its shadow) $h\pm \nu=\Delta$ (where $h=\frac{d}{2}$).   But there are additional poles corresponding to the ``double trace" operator $2\D_{\phi}$ \cite{note1}.
 
In Mellin space we can write the corresponding spectral representation as 
\be\label{sunitrymell}
M_{\Delta,\ell}^{(s)}(s,t) = \int_{-i\infty}^{i\infty} d\nu\, \mu_{\Delta,\ell}(\nu)
\Omega_{\nu, \ell}^{(s)}(s)P^{(s)}_{\nu, \ell}(s,t)\,.
\ee
The conformal partial waves go over to a set of so-called Mack polynomials $P^{(s)}_{\nu, \ell}(s,t)$ of degree
 $\ell$ in $(s,t)$ \cite{mack,dolanosborn,regge}. We also have an additional factor
\be\label{oms}
\Omega_{\nu, \ell}^{(s)}(s) = \frac{\G(\frac{h+\nu-\ell}{2}-s)\G(\frac{h-\nu-\ell}{2}-s)}{\G^2(\Delta_{\phi}-s)}\,.
\ee  

We now pick out the spurious poles as in (\ref{polexp}). Firstly, (\ref{oms}) has a denominator piece which cancels against the $\G^2(\Delta_{\phi}-s)$ in $\rho_{\Delta_{\phi}}(s,t)$. 
Secondly, note that the poles at $2\D_{\phi}-h+\ell-\nu=0$ in the numerator of (\ref{specunitry}) give rise, upon doing the $\nu$ integral to the required single and double spurious poles at $s=\D_\phi$.  
Finally, we observe that the Mack Polynomial defines, through
\be\label{mackres}
Q^{\Delta}_{\ell,0}(t)= \frac{4^\ell }{(\Delta-1)_{\ell}(2h-\Delta-1)_{\ell}}P^{(s)}_{\Delta-h,\ell}(s=\frac{\Delta-\ell}{2},t) \,,
\ee
a single variable orthogonal polynomial (labelled by $\ell$ and explicitly expressible in terms of hypergeometric functions) known as a continuous Hahn Polynomial -- see \cite{suppl,regge} for details. They are the analogue of Legendre Polynomials in a partial wave expansion. This is a particularly nice feature of the Mellin expansion since it gives us a way to decompose the residues in (\ref{polexp}) in a natural basis and impose the condition of vanishing on the coefficients term by term. Moreover, what we have just seen is that in the $s$-channel, a field of a given spin $\ell$ only contributes to the $Q^{\D}_{\ell,0}(t)$ with the same $\ell$. Thus we can write this contribution to (\ref{qsum}) as 
\be\label{qs}
q^{(a,s)}_{\D, \ell}(t)= q^{(a,s)}_{\D, \ell}Q^{2\D_\phi+\ell}_{\ell,0}(t)
\ee
with  $q^{(2,s)}_{\D, \ell},q^{(1,s)}_{\D, \ell}$ being the coefficients of the constant and $(s-\D_\phi)$ term from 
\be
q_{\D,\ell}^{(s)}(s)=-\frac{4^{1-\ell}\G(\D_\phi+s+\ell-h)^2}{(\ell+2s-\D)(\ell+2s+\D-2h)\G(2s+\ell-h)}\,,
\ee
under a Taylor expansion around $s=\D_\phi$.
%\be\label{q1s}
%q^{(2,s)}_{\D, \ell}=-\frac{4^{1-\ell}\G(2\D_\phi+\ell-h)}{(\ell-\D+2\D_\phi)(\ell+\D+2\D_\phi-2h)}
%\ee
%and
%\be\label{q2s}
%q^{(1,s)}_{\D, \ell}=\frac{4^{2-\ell}\G(2\D_\phi+\ell-h+1)}{(\ell-\D+2\D_\phi)^2(\ell+\D+2\D_\phi-2h)^2}.
%\ee

This was for the $s$-channel but we can add in the $t,u$-channels easily by appropriate exchange of $u,v$ variables which translates into exchanging the Mellin variables with some shifts: 
\begin{eqnarray}\label{shifts}
t-channel&:& s\rightarrow t+\D_\phi, t\rightarrow s-\D_\phi \,,\nonumber\\ u-channel&:& s\rightarrow \D_\phi-s-t, t\rightarrow t\,.
\end{eqnarray}
%\begin{align}\label{totmell}
%\begin{split}
%{\cal M}(s,t)= &   \sum_{\Delta, \ell} c_{\D, \ell} (M_{\Delta,\ell}^{(s)}(s,t)+M_{\Delta,\ell}^{(t)}(s,t)
%+M_{\Delta,\ell}^{(t)}(s,t)) \\
%= \sum_{\Delta, \ell} c_{\D, \ell} & \int_{-i\infty}^{i\infty} d\nu\, 
%( \mu^{(s)}_{\Delta,\ell}(\nu)\Omega_{\nu, \ell}^{(s)}(s)P^{(s)}_{\nu, \ell}(s,t)
%+\mu^{(t)}_{\Delta,\ell}(\nu)\Omega_{\nu, \ell'}^{(t)}(t)
%P^{(t)}_{\nu, \ell'}(s-\frac{1}{2}(\D_3+\D_4),t+\frac{1}{2}(\D_2+\D_3))  \\
%+& \mu^{(u)}_{\Delta,\ell}(\nu) \Omega_{\nu, \ell'}^{(u)}(s+t) P^{(u)}_{\nu, \ell'}(s-\frac{1}{2}(\D_1+\D_4),t ) ) \, 
%\end{split}
%\end{align}
We need to extract the corresponding contributions $q^{(a,t)}_{\D, \ell'}(t)$ to the residues and decompose them
in an expansion in the same orthogonal basis of $Q^{2\D_\phi+\ell}_{\ell,0}(t)$. Now, however, a spin $\ell'$ exchange in these channels will give a contribution in all partial waves. With the change of variables explained above, we have 
\be
M_{\Delta,\ell'}^{(t)}(s,t) =M_{\Delta,\ell'}^{(s)}(t+\D_\phi,s-\D_\phi) \,.
\ee
Now the $\rho_{\Delta_{\phi}}(s,t)$ gives rise to the spurious poles and thus we just need to evaluate $M_{\Delta,\ell'}^{(t)}(s,t)$ and its first order expansion around $s=\D_\phi$ to obtain $q^{(a,t)}_{\D, \ell}(t)$. Furthermore, the individual contributions to the $Q^{2\D_\phi+\ell}_{\ell,0}(t)$ expansion can be picked out using their orthogonality properties. The end results for $q^{(2,t)}_{\D, \ell},q^{(1,t)}_{\D, \ell}$ are obtained, as before, by Taylor expanding
\begin{align}\label{q2t}
c_{\D,\ell}q^{(t)}_{\D, \ell}(s) &=\k_\ell(s)^{-1}\sum_{\ell'}c_{\D,\ell'} \int dt d\n \G^2(s+t)\G^2(-t) 
\nonumber \\&
\!\!\!\!\!\!\!\!\!\!\times \mu_{\Delta,\ell'}(\nu) \Omega_{\nu, \ell'}^{(t)}(t) P^{(t)}_{\nu, \ell'}(s-\D_\phi,t+\D_\phi) Q^{2s+\ell}_{\ell,0}(t)  \,,
\end{align}
around $s=\D_\phi$.
Here $\kappa_\ell(s)$ is a normalization factor \cite{suppl} and $P^{(t)}_{\nu,\ell'}(s,t)=P^{(s)}_{\nu,\ell'}(t,s)$.
It can be shown straightforwardly, using the properties of the continuous Hahn polynomials that the $u$-channel gives an identical contribution i.e. $q^{(a,u)}_{\D, \ell} =q^{(a,t)}_{\D, \ell}$. 

The sum over the physical spectrum also includes the identity operator ($\D=\ell=0$). It will be convenient to separate out this piece.  It gives a position space contribution to ${\cal A}(u,v)$ which is $(1+(\frac{u}{v})^{\D_\phi}+u^{\D_\phi})$.  We will take the corresponding Mellin amplitude to be given by the poles that reproduce this power law behaviour. Thus
\be
M_{\D=0,\ell=0}(s,t) =\rho_{\D_\phi}(s,t)^{-1}(\frac{1}{st}+{\rm crossed})\,,
\ee
where the crossed channels are obtained from the s-channel using eq.(\ref{shifts}).
In this case only the $t,u$-channels contribute to a spurious single  pole at $s=\D_\phi$. The contribution to 
$Q^{2\D_\phi+\ell}_{\ell,0}(t)$ can be evaluated by using the above amplitude and orthogonality.
The answer is
\be
q^{(1,t)}_{\D=0, \ell}=q^{(1,u)}_{\D=0, \ell}= -   \k_\ell(\D_\phi)^{-1}Q^{2\D_\phi+\ell}_{\ell,0}(0)
 \ee
 
Thus the simplest set of bootstrap equations \cite{desc} in Mellin space read as 
\be\label{btstrpfin}
\!\!\!\sum_{\D \neq 0,\ell} \!\!\!c_{\Delta,\ell}(q^{(2,s)}_{\D, \ell} +2q^{(2,t)}_{\D, \ell})=0=2q^{(1,t)}_{\D=0, \ell}+\!\!\sum_{\D \neq 0,\ell}\!\! \!\!c_{\Delta,\ell}(q^{(1,s)}_{\D, \ell}+2q^{(1,t)}_{\D, \ell})\,.
\ee
We have an infinite number of equations, one for each $\ell$. The first term corresponds to the vanishing of the log term and the second to the spurious power law piece in position space. Typically the latter constraint determines anomalous dimensions and the former, OPE coefficients.

\section{Results}
\label{sec:results}
The scalar $\phi^4$ theory in $d<4$ has an interacting fixed point in the IR known as the Wilson-Fisher fixed point. 
This fixed point is accessible perturbatively in an $\e$ expansion where $d=(4-\e)$. The anomalous dimension of $\phi$ and $\phi^2$ are known upto $\e^5$ order e.g. \cite{kleinert,guida} while for the higher spin operators, $J^{(\ell)}$ the result is known to $\e^4$ order \cite{gracey, analyt}. However, Feynman diagram computations for OPE coefficients for the stress tensor exchange, have only been carried out to a couple of low orders in $\e$. Here we will apply the above bootstrap procedure to the four point function of $\phi$. 
We will also assume the existence of a unique stress tensor with $(\D=d, \ell=2)$ as the lowest member of a tower of twist two primaries $J^{(\ell)}$ of even spin $\ell$.  By demanding the cancellation of the spurious terms \cite{suppl} we find
$
\Delta_\phi =1-\frac{\epsilon }{2}+\frac{1}{108} \epsilon^2+\frac{109}{11664} \epsilon^3+O(\e^4)$, $\D_{\phi^2}=2-\frac{2}{3}\e+\frac{19}{162}\e^2+O(\e^3)$ which reproduce 3-loop Feynman diagram results. What is nontrivial to obtain diagrammatically is the OPE coefficient \cite{opecomm}, with $\phi^2$ exchange, 
where we have a prediction at $O(\e^2)$.
\be
\frac{C_{0}}{C_0^{free}}=1-\frac{1}{3}\e-\frac{17}{81}\e^2+O(\e^3)\, .
\ee
Here we have normalized the result with the free theory OPE coefficient and have written $ {\mathcal C}^2_{\phi\phi\phi^2}=C_0$. The $O(\e^2)$ result yields $C_0/C_0^{free}\approx 0.457$ on setting $\e=1$ as compared to $0.553$ from numerics \cite{mostprecise}.
One can go onto studying the sector with higher spin currents $J^{(\ell)}$ in an analogous fashion. We simply state the results (details to appear in \cite{long})
\be
\D_\ell=d-2+\ell+\left(1-\frac{6}{\ell (\ell +1)}\right)\frac{\epsilon^2}{54}+\d^{(3)}_{\ell}\e^3+O(\e^4).
\ee
We recover the known $O(\e^2)$  \cite{wilsonkogut} and  $O(\e^3)$ results  \cite{gracey} 
\be\label{ecub}
\!\! \d^{(3)}_{\ell}=\frac{373\ell^2-384\ell-324+109 \ell ^3(\ell+2)-432\ell(\ell+1)H_{\ell}}{5832 \ell ^2 (\ell+1 )^2}
\ee
where $H_n$ denotes the Harmonic number. We also have
%As a check, this nontrivial function vanishes for the stress-tensor, $\ell=2$. Finally, we also obtain the OPE coefficient
%\begin{widetext}
\begin{eqnarray}\label{cell}
\frac{C_{\ell}}{C_{\ell}^{free}}&=& 1+\frac{\epsilon ^2}{54\ell (\ell+1)}\left[6(\ell+1)^{-1}+2 (\ell^2+\ell-3)H_\ell \right. \nonumber\\&&~~~~~~~~~~~ - (\ell-2)(\ell+3)H_{2\ell}\big{]}+ {\mathcal C}_\ell^{(3)}\e^3\,,
\end{eqnarray}
%\end{widetext}
where $C_{\Delta_\ell,\ell}=C_\ell$.
This is a completely new result.
The $O(\e^3)$ term can also be calculated case by case for any given spin \cite{long}.
In particular, this implies that the central charge  $c_T=\frac{d^2 \Delta_\phi ^2}{(d-1)^2 C_{2}}$ is given by
\be
\frac{c_T}{c_{free}}=1-\frac{5 \epsilon ^2}{324}-\frac{233 \epsilon ^3}{8748}+O(\e^4)\,.
\ee
While the $O(\e^2)$ is known e.g., \cite{petkou}, the $O(\e^3)$ order is new. If we put $\e=1$ and compare with the 3d Ising model numerical result, $c_T/c_{free}=0.946534(11)$, from bootstrap \cite{3dising}, we get with $\epsilon=1$,  
\be
c_T/c_{free}\approx 0.957933\,,
\ee
which is a better estimate than what one gets from only the $O(\e^2)$ part ($\sim 0.98$). 
%{\bf AS: not sure if we should leave the following in since giombi has made a similar stmt in his paper: probably we should just remove it and quote the $\e=1$ values for the ope coeffs for future numerics} Another example is the comparison between the $\ell=4$ conformal dimension which numerics gives as 5.0208(12) \cite{ell4}. The $O(\e^2)$ result yields 5.01296 while the $O(\e^3)$ improves it to 5.02495.  {\bf AS: I suggest we change to: 
Our $O(\e^3)$ explicit results for OPE coefficients gives ${C_{4}}/{C_{4}^{free}}=1.07872$ for $\e=1$. Numerical bootstrap results for this coefficient are scarce and, as yet, with undetermined errors \cite{zohar2}. Using the results in  \cite{zohar2}, numerics yield 1.11345 \cite{commn}. 
In fact, 
the $O(\e^2)$ results (\ref{cell}) as well as the $O(\e^3)$ results \cite{long}, show that, as a function of $\ell$, ${C_{\ell}}/{C_{\ell}^{free}}$ exhibits a minimum at $\ell=4$. It will be interesting to see whether numerical bootstrap finds a similar feature.

Denoting the anomalous dimension of the higher spin $J^{(\ell)}$'s  by $\gamma_\ell$ and that of $\phi$ by $\gamma_\phi$, using our methods, it is also possible to derive the following universal form for $\gamma_\ell$ in the limit $\ell\gg 1$ for weakly coupled theories (with a small twist gap and coupling $g\ll 1$) in $d$-dimensions:
\be \label{weakg}
\gamma_\ell-2\gamma_\phi=\frac{\sum_{p=0}^\infty \alpha_p(g)(\log \ell)^p}{\ell^{d-2}}
\,,
\ee
whose form agrees with \cite{alday} but our method gives explicitly for $d=4-\e$ (where $g=\e$)
\be \label{ap2}
\alpha _p(\e) =-\frac{\e^{2+p}}{9p!}\left(\frac{2}{3}\right)^p+O\left(\e^{3+p}\right)\,,
\ee
 which can be cross-checked for $p=0,1$ using (\ref{ecub}). The general $p$ formula is a prediction. Notice that, plugging the leading order $\alpha_p$ into (\ref{weakg}) resums into $-\e^2/(9\ell^{2-2\e/3})\approx -\e^2/(9\ell^{\Delta_{\phi^2}})= -\e^2/(9\ell^{\tau_{\phi^2}})$ where $\tau_{O}$ is the twist of the operator $O$. This spin-dependence and the coefficient are in agreement with what would follow from \cite{dlce}. A similar analysis can be done for any weakly coupled theory \cite{long}.

Our approach can also be used to get the leading anomalous dimensions for the $\phi^3$ theory in 6-dimensions as well as the $\phi^6$ theory in 3-dimensions \cite{long} as well as results for $O(N)$ \cite{kds} both at fixed $d$, large $N$ as well as in the $\e$-expansion. It will also be interesting to extend our techniques to the theories being investigated in \cite{epsmotiv}.

\section{Outlook}
\label{sec:outlook}

The new approach to bootstrap that we have outlined worked remarkably well for the Wilson-Fisher fixed point, reproducing analytically known results and producing new results for OPE coefficients. In contrast to the complexity of higher loop Feynman diagram computations, with all their divergences and regularisations, our method yields finite, scheme independent physical results with relative ease of calculation. The main reason for this efficacy is that in all the results we have discussed, the crossed channels involved at most the identity operator and $\phi^2$ with other operators contributing only at higher orders. This simplification does not occur in the conventional approach to bootstrap \cite{rrtv} -- there one typically needs to sum over an infinite set of operators \cite{dlce}, even to produce results at leading order in epsilon at large spin \cite{sensinha}. At higher orders this phenomenon is unlikely to persist and we will have to perhaps make use of the additional spurious poles \cite{desc}. It is likely, however, that in the presence of a small parameter $(\e, \frac{1}{\ell}, \frac{1}{N}, \ldots)$ we can always obtain the leading results analytically. 
%This coincidence is unlikely to persist at higher orders and for other general theories and we will very likely need to make use of the spurious poles associated with descendant states, namely $s=\D_\phi+n$ to be able to systematically continue the expansion. Of course, rather than use an epsilon-expansion, one could use our equations to expand in other parameters.

The conceptual suggestiveness of the present approach lies in the $AdS_{d+1}$ Feynman diagram-like expansion. When combined with the Mellin representation, this holds out the tantalising possibility of deciphering a dual string theory interpretation for CFTs, at least where a large $N$ limit exists.  
%One of the main advantages of using Mellin space is that it opens up the possibility of exploring various limits of $s,t$ for example the Regge limit \cite{long} and systematically studying universal properties of any CFT spectrum. For instance, one could ask if the large conformal dimension sector of CFTs allow for a systematic solution using the inverse conformal dimension as an expansion parameter. Blah.

\vskip 5mm

{\bf Acknowledgments} : 
We would especially like to thank J. Penedones for collaboration during the initial stages of this work and for discussions. We acknowledge useful discussions with S. Giombi, I. Klebanov, Z. Komargodski, J. Maldacena, G. Mandal, S. Minwalla, H. Osborn, S. Raju and especially S. Rychkov. R.G. acknowledges the support of the J. C. Bose fellowship of the DST. A.S. acknowledges support from a DST SwarnaJayanti Fellowship Award DST/SJF/PSA-01/2013-14. More generally, this work would not have been possible without the unstinting support for the basic sciences by the people of India.

\clearpage

\onecolumngrid
\vskip 2cm
\begin{center}
{\bf {\large Supplementary material}}
\end{center}
\appendix

\section{Mack polynomials}
The explicit expression for the Mack polynomials--see e.g.\cite{mack,dolanosborn,regge}-- $P^{(s)}_{\nu,\ell}(s,t)$ in our conventions is given by:
\begin{align}\label{sumt}
\begin{split}
& P^{(s)}_{\nu,\ell}(s,t)=\widetilde{\sum}\frac{\Gamma^2(\l_1)\Gamma^2(\bar\l_1)(\l_2-s)_k(\bar{\l}_2-s)_k(s+t)_\b(s+t)_\a(-t)_{m-\a}(-t)_{\ell-2k-m-\b}}{\prod_i\G(l_i)}\,,\\
& \text{where \ \  }\widetilde{\sum}\equiv 2^{-\ell}\ell! \sum_{k=0}^{[\frac{\ell}{2}]}\sum_{m=0}^{\ell-2k}\sum_{\a=0}^m\sum_{\b=0}^{\ell-2k-m}(-1)^{\ell-k-\a-\b}\frac{(\ell+h-1)_{-k}}{k!\a!\b! (m-\a)!(\ell-2k-m-\b)!}\,.
\end{split}
\end{align}
Here $(p)_q=\Gamma(p+q)/\Gamma(p)$ is the Pochhammer symbol and
\be
\l_1=\frac{h+\nu+\ell}{2} \,, \ \ \bar{\l}_1=\frac{h-\nu+\ell}{2}\,, \ \ \l_2=\frac{h+\nu-\ell}{2}\text{ \ \  and \ \  }\bar{\l}_2=\frac{h-\nu-\ell}{2}\,,
\ee
and the $l_i$-s are given by,
\be\label{li}
l_1=\l_2+\ell-k-m+\a-\b\,,\ l_2=\l_2+k+m-\a+\b\,, \ l_3=\bar{\l}_2+k+m\,,\ l_4=\bar{\l}_2+\ell-k-m\,.
\ee

\section{Continuous Hahn Polynomials}\label{A}
The continuous Hahn  polynomials $Q^{\D}_{\ell,0}(t)$ defined via (eq.10 in the main text) can be shown to be equal to \cite{regge}
\be\label{Qdefn}
Q^{2s+\ell}_{\ell,0}(t)=\frac{2^\ell ((s)_\ell)^2}{(2s+\ell-1)_\ell}\ {}_3F_2\bigg[\begin{matrix} -\ell,2s+\ell-1,s+t\\
s, s
\end{matrix};1\bigg]\,.
\ee
The orthonormality condition for these  $Q_{\ell,0}$ polynomials is given by \cite{AAR}
\be
\frac{1}{2\pi i}\int_{-i\infty}^{i\infty}\G^2(s+t) \G^2(-t) Q^{2s+\ell}_{\ell,0}(t)Q^{2s+\ell '}_{\ell',0}(t)=\kappa_{\ell}(s)\d_{\ell,\ell'}\,,
\ee
where,
\be
\kappa_{\ell}(s)=\frac{4^\ell \ell! \G^4(\ell+s)\G(2s+\ell-1)}{\G(2s+2\ell)\G(2s+2\ell-1)}\,.
\ee

\section{Calculational details}
In the $\e$ expansion, we let 
\begin{eqnarray}
\D_\phi&=&1+\d_\phi^{(1)} \e+\d_\phi^{(2)} \e^2+\ldots \,,\\ 
\D_{\phi^2}&=&2+\d_{\phi^2}^{(1)} \e+\d_{\phi^2}^{(2)} \e^2+\ldots\,,\\
\D_{\ell}&=&d-2+\ell+\d_\ell^{(1)} \e+\d_\ell^{(2)} \e^2+\ldots\,,
\end{eqnarray} and the coefficient 
\be
C_{\D_\ell,\ell} \equiv C_\ell = C_{\ell}^{(0)}+C_{\ell}^{(1)}\e+C_{\ell}^{(2)}\e^2+\ldots \,.
\ee
$C_0$ will denote the coefficient with the $\phi^2$. The OPE coefficients have been normalized so that in position space, in the $u\rightarrow 0,v\rightarrow 1$ limit we get $C_{\Delta_\ell,\ell}u^{(\Delta-\ell)/2}(1-v)^\ell$ as the leading term for ${\cal A}(u,v)$. We give results in the main text by dividing by the free theory result in this convention so that the final answer is convention independent.

What helps us to solve the bootstrap equation to low orders in $\e$ is the fact that in the crossed-channels all nontrivial operators start contributing to $q^{(a,t)}_{\D, \ell>0}$ only from $O(\e^2)$ \cite{long}. Thus to $O(\e)$ we must have $q^{(1,s)}_{\D=d, \ell=2}+2q^{(1,t)}_{\D=0, \ell=2}=0$ and $q^{(2,s)}_{\D=d, \ell=2}=0$. This determines
\be
\d^{(1)}_\phi=-\frac{1}{2}\,,\hspace{1cm}C_{2}^{(0)}=\frac{1}{3}\hspace{1cm}\text{and}\hspace{1cm}C_{2}^{(1)}=-\frac{11}{36}\ .
\ee
We can now bootstrap to higher orders by looking at the $\ell=0$ term. There is a contribution only from $\phi^2$ in the $t$-channel. The bootstrap equations are now $q^{(2,s)}_{\D_{\phi^2}, \ell=0}+2q^{(2,t)}_{\D_{\phi^2}, \ell=0}=0$ and $q^{(1,s)}_{\D_{\phi^2}, \ell=0}+2q^{(1,t)}_{\D=0, \ell=0}+2q^{(1,s)}_{\D_{\phi^2}, \ell=0}=0$. This determines
\be
C^{(0)}_{0}=2\,,\hspace{1cm}C^{(1)}_{0}=-\frac{2}{3}\hspace{1cm} \text{and}\hspace{1cm}\d^{(1)}_{\phi^2}=-\frac{2}{3}\,.
\ee
Plugging these back into the $\ell=2$ equation enables one to solve to the next order and iterating once more with $\ell=0$ gives  the results in the main text.

To derive the large spin results for the weakly coupled theories, we need the large $\ell$ limit of $Q_{\ell,0}$'s which is given by
\be
Q_{\ell,0}^{2s+\ell}(t) =\frac{2^\ell\ell ^{-s-t} \Gamma (s+\ell )^2 \Gamma (s-t+\ell-1 )}{\Gamma (-t)^2 \Gamma (2 s+2 \ell -1)}\,.
\ee
The s-channel has a single spin-$\ell$ operator with dimension $\Delta=d-2+\ell+\gamma_\ell$ with the corresponding orthogonal polynomial taking the above form. The integral in (eq.15) now simplifies since we are interested only in the leading order result in large spin. The leading exchange in the crossed channel for $d\leq 4$ can be shown to be a scalar with dimension $d-2+\gamma_0$. For $d>4$ the leading spin dependence arises due to a scalar operator with leading dimension 2 which is the same as that of the lagrange multiplier field.  In a weakly coupled theory, other operators which involve higher powers of the elementary field turn out to contribute at a subleading order.
It can be checked that in order to solve both the constraints at leading order in the coupling arising from the cancellation of the double and single spurious poles, we need the large spin OPE coefficient to be that in the mean field theory and the large spin anomalous dimensions to satisfy (for $d\leq 4$)
\be
\gamma_\ell-2\gamma_\phi=\frac{\sum_{p=0}^\infty \alpha_p(g)(\log \ell)^p}{\ell^{d-2}}
\,,
\ee
with
\be
\alpha _p=- \frac{C_{d-2,0}}{C^{free}_{\ell=0}} g^{2+p}\frac{(-\delta^{(1)}_{0})^p}{{p!}}   \left(\delta^{(1)}_{0}-2 \delta^{(1)}_{\phi }\right)^2 \Gamma (d-2) +O\left(g^{3+p}\right)\,.
\ee
For the $\e$ expansion, where $ \delta_\phi^{(1)}=-1/2$ and $\delta_{\phi^2}^{(1)}=-2/3$ we obtain (eq.26) in the main text. One can check that when other operators start contributing they will do so at a subleading order in $\e$ to each power of $\log \ell$.

\section{Convergence properties of eq (3)}
We will briefly discuss the convergence properties of eq. (3) in the main text. In the existing numerical approach to the bootstrap, it is known that the corresponding sum over the spectrum is absolutely convergent \cite{rychconv}. In Eq.(1) or (3), there is a sum over $\D$ as well as $\ell$. We make preliminary comments on the convergence of both sums in turn. In considering the sum over $\D$ in Eq. (1) or (3), we recall that each Witten diagram contribution can be decomposed in terms of  conformal blocks in the schematic form (see for e.g. Eq. (4.16) of \cite{perl})
\be\label{wittdecomp}
W^{(s)}_{\D,\ell}(u,v) = N_{\D,\ell}G^{(s)}_{\D,\ell}(u,v)+\sum_n N^{\D}_{n,\ell} G^{(s)}_{2\D_\phi +2n+\ell}(u,v).
\ee
Here the second term is the sum over double trace operators of the schematic form $\phi(\partial^2)^n\partial^{\ell} \phi$. The coefficients $N^{\D}_{n,\ell}$ are known explicitly. 
These give rise to the spurious poles.
We can write similar expansions in the $t,u$ channels. 
For the physical pole (the first term) we know that $c_{\D,\ell}N_{\D,\ell} =C_{\D,\ell}$ where the latter is the square of the OPE coefficient. It is the convergence of this sum which has been studied in detail in \cite{rychconv}. On the other hand, each spurious pole (labelled by $n$) gets a contribution from every $\D$ of the form (in the $s-$channel) $\sum_{\D}c_{\D,\ell}N^{\D}_{n,\ell}$. While we postpone a careful analysis to later, we can see that since $N^{\D}_{n,\ell}$ has a similar behaviour to  $N_{\D,\ell}$, this sum is similar to the sum over OPE coefficients $C_{\D,\ell}$. We know from \cite{rychconv}, that the OPE coefficients are exponentially suppressed in a way such that even after convolution with the spectral density, it leads to an absolutely convergent series. We therefore might expect a similar absolute convergence of the contributions from the $\D$ sum to each spurious pole. 

We next turn to the sum over spin $\ell$ which is relatively more under control. Here we can perform the d'Alembert convergence test using the summands in eq.(3) concretely in the $\epsilon$ expansion. We will work with
\be\label{summand}
r_\ell=\sum_{\Delta}(M^{(s)}_{\Delta,\ell}(\D_\phi,t)+M^{(t)}_{\Delta,\ell}(\D_\phi,t)+M^{(u)}_{\Delta,\ell}(\D_\phi,t))\,,
\ee
and consider $\beta_\ell=\displaystyle|\frac{r_{\ell+2}}{r_\ell}|$.

\begin{figure}[t!]
  \centering
  \begin{tabular}{cc}
        \includegraphics[width=0.5\textwidth]{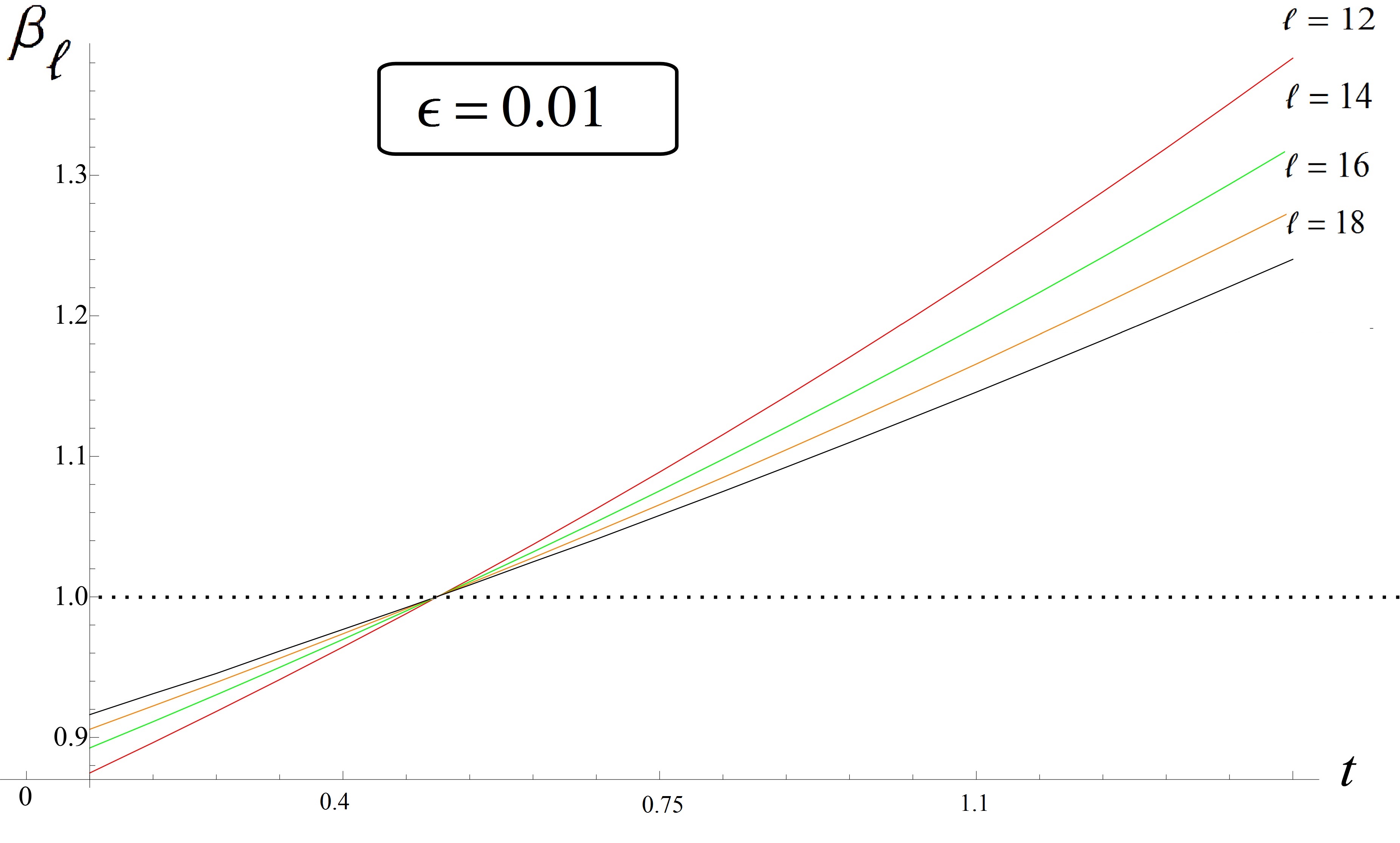} & \includegraphics[width=0.5\textwidth]{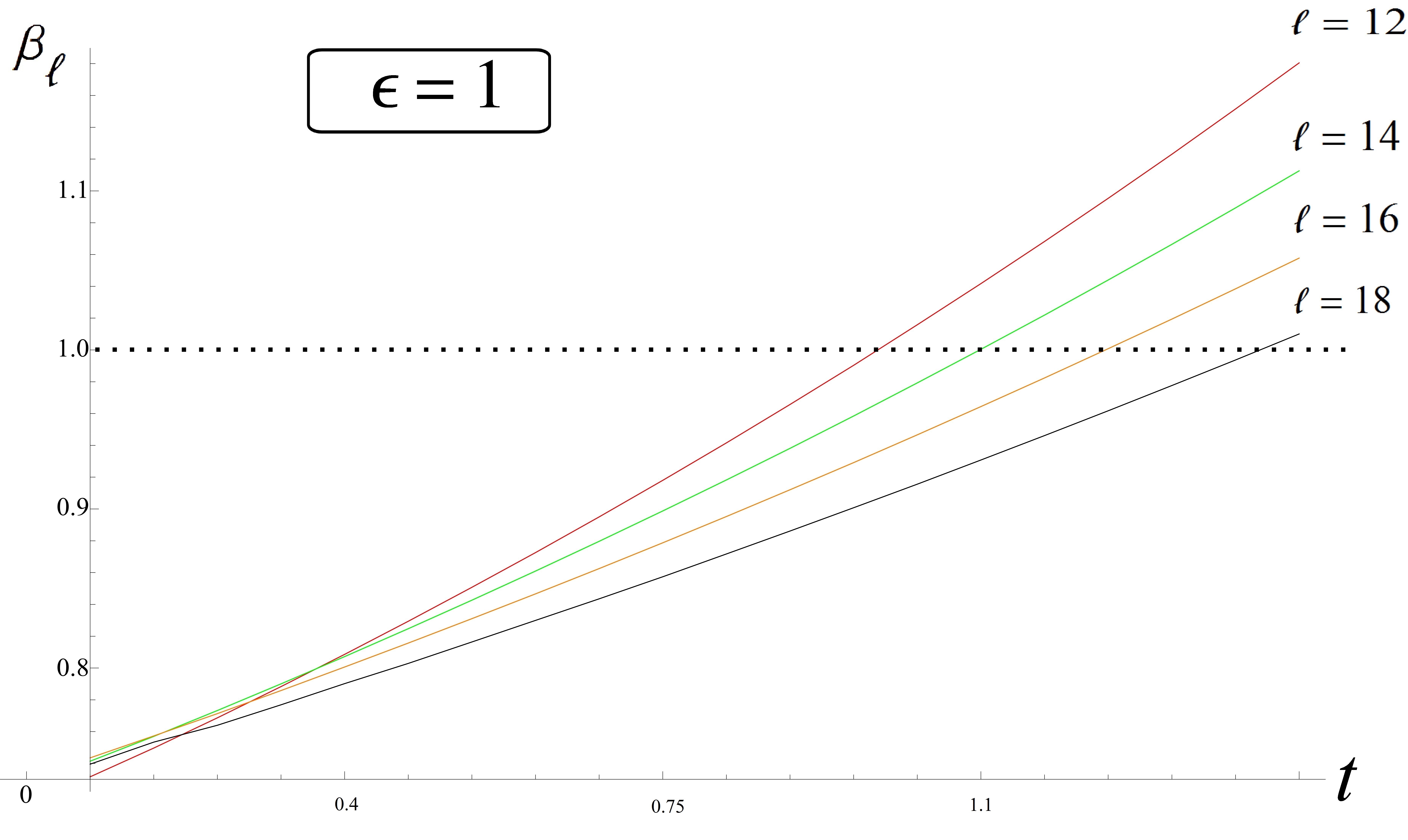}
\end{tabular}
    \caption{Plot of $\beta_\ell$ as a function of $t$ for different values of $\epsilon$. }
\end{figure}

\begin{figure}[b!]
  \centering
        \includegraphics[width=0.5\textwidth]{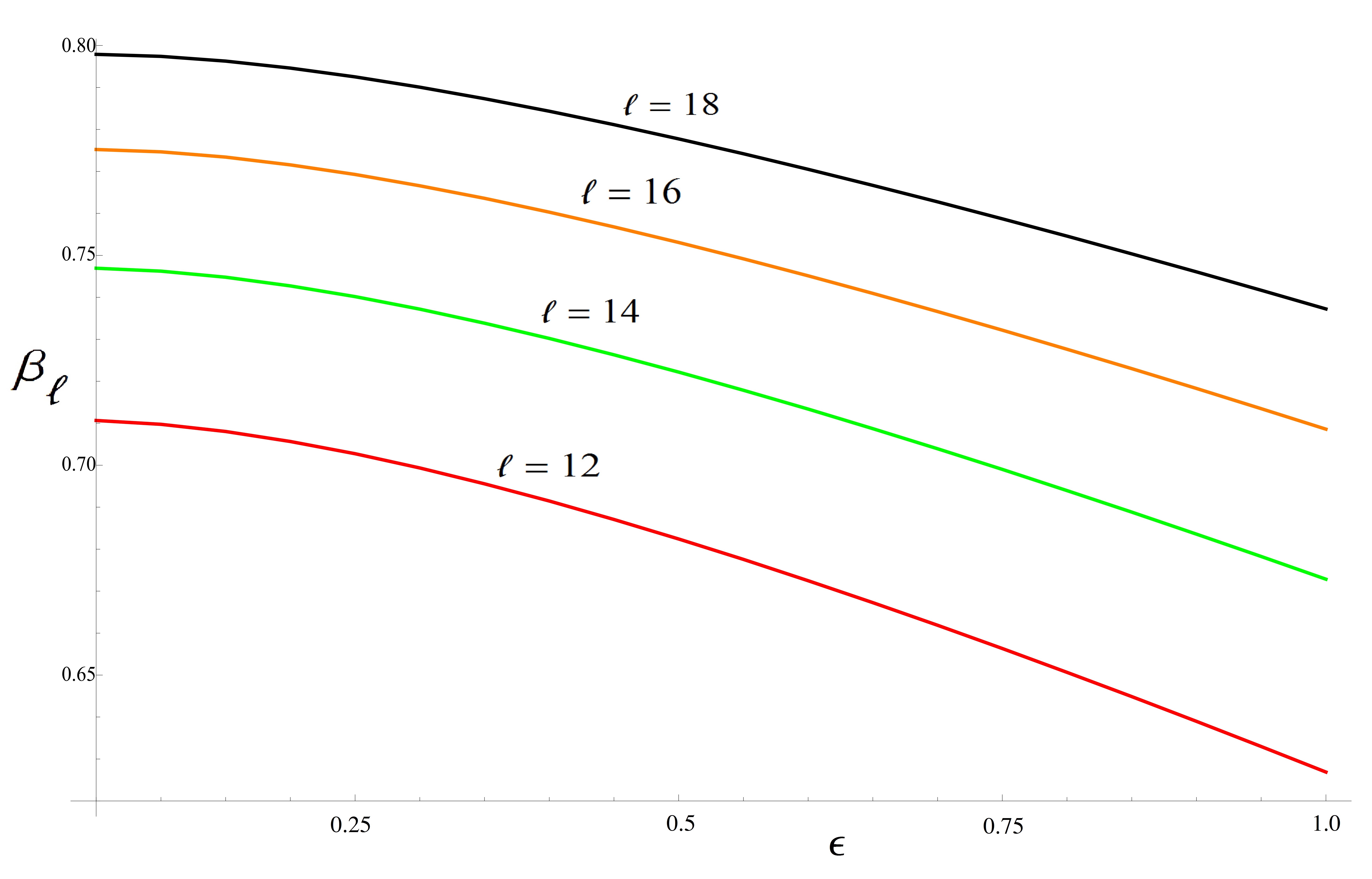}
    \caption{Crossed channel sum over spins at $O(\e^4)$ is absolutely convergent. Here $r_\ell=c_{\Delta,\ell} q_{\Delta,\ell}^{(2,t)}$ and  $\beta_\ell=|\frac{r_{\ell+2}}{r_\ell}|$. In the $\e$-expansion only the higher spin currents $J^{(\ell)}$ contribute at this order. Assuming this continues to hold for $\e\sim O(1)$ and using our $O(\e^3)$ results, we have obtained the above plot.  }
\end{figure}

Specifically we will perform  numerical checks to provide evidence that there exist ranges of $t$ where the sum is absolutely convergent, in other words $\beta_\ell<1$ as $\ell\gg 1$--this is shown in figure 1. We will take the $O(\e^3)$ results obtained in the $\e$-expansion and plot $\beta_\ell$ as a function of $t$.  The analogous analysis using the newly obtained results for a large class of operators in the 3d-Ising model \cite{DSD} is similar. In deriving the $\epsilon$-expansion $O(\e^3)$ results we used the fact that in the crossed channels only the scalar operator contributed upto this order but the higher spin operators contributed at $O(\e^4)$. In figure 2 we show that the sum over the spin in the crossed channel is also absolutely convergent and hence we are justified in dropping these terms. We can in fact show that each term in the $O(\e^4)$ summand (which correspond to higher spin current $J^{(\ell)}$ exchange) is positive and is less than $\displaystyle \frac{\e^4}{81 \ell^4}$. Using this result it is easy to conclude that the sum is absolutely convergent.

 Another evidence that our $O(\e^3)$ results are correct is provided in \cite{PDAKAS} where we derive the $O(\e^3)$ $c_T$ using Feynman diagram based arguments and find exact agreement with our bootstrap results.
We will leave a rigorous analysis along the lines of \cite{rychconv} for future work to identify the appropriate set of conditions which will enable a systematic study of numerics using our equations.


\begin{thebibliography}{99}
\bibitem{commconv} We discuss the convergence properties of this expansion in \cite{suppl}. In the previous approach following \cite{rrtv}, the sum over the spectrum is known to be absolutely convergent \cite{rychconv}.
\bibitem{poswit}
There is a large literature on calculations of Witten diagrams in position space starting with \cite{fmmr}.
\bibitem{compl} In an interacting superconformal theory, there could be special operators (``chiral primaries") such that there are are physical operators in the spectrum with dimension $2\D_\phi$. In such cases, we would have to consider mixed correlators or other ways to sidestep this.
\bibitem{note1}
Note: there are further poles from the Pochammer terms $(h\pm \nu-1)_\ell$, and hence one actually needs to add additional terms to cancel these contributions. We will not include the contributions of these additional poles since the additional terms do not contribute to the quantities of interest to us. Further, the integrals over the spectral parameter can in fact be done exactly, leading to a huge simplification in the expressions for the blocks \cite{rgas}.
\bibitem{desc} We have considered here the simplest set of constraints from the poles at $s=\D_\phi$. There are subsidiary spurious poles at $s= \D_\phi+m$ for $(m=1,2\ldots$).
\bibitem{analyt} 
Non-Feynman diagram analytic approaches to computing anomalous dimensions have reproduced the leading nontrivial anomalous dimensions of $\phi$ to $O(\e^2)$, $\phi^2$ to $O(\e)$ \cite{rychkovtan} and $J^{\ell}$  to O($\e^2$) \cite{skvort, giombi}. Ref \cite{sensinha} uses the consistency of OPE with the dispersion relation as advocated in \cite{Polyakov} to derive the $O(\e^2)$ anomalous dimensions for $\phi,\phi^2$.
\bibitem{opecomm} In this section the OPE coefficients (squared) $C_{\Delta_\ell,\ell}$'s are normalized as explained in \cite{suppl}.
\bibitem{commn} $C_{\ell\neq 0}^{free}=2^{-2\ell+2}$ in \cite{zohar2}. If we normalized our blocks in the same way as \cite{zohar2}, $\sqrt{C_{4}}\approx 0.1337$ compared to $0.1319$ in \cite{zohar2}. The $O(\e^3)$ was very important to get this agreement.
\end{thebibliography}
\end{document}